\documentclass[11pt]{article}
\usepackage[leqno]{amsmath}
\usepackage{amsthm}
\usepackage{amsfonts}
\usepackage{amssymb}
\usepackage{eucal}

\theoremstyle{plain}
  \newtheorem{theorem}{Theorem}
  \newtheorem{proposition}[theorem]{Proposition}
  \newtheorem{lemma}[theorem]{Lemma}

\theoremstyle{definition}
  \newtheorem{example}[theorem]{Example}
  \newtheorem{remark}[theorem]{Remark}

\newcommand{\R}{{\mathbb R}}
\newcommand{\C}{{\mathbb C}}

\newcommand{\leftbracket}{\langle}
\newcommand{\rightbracket}{\rangle}

\newcommand{\g}{{\frak g}}
\newcommand{\so}{{\frak so}}
\DeclareMathOperator{\Ad}{Ad}
\DeclareMathOperator{\trace}{Tr} \DeclareMathOperator{\End}{End}  \DeclareMathOperator{\diam}{diam}

\begin{document}

\title{Diameters of Homogeneous Spaces}

\author{Michael H. Freedman$\footnote{Microsoft Research,One Microsoft Way, Redmond, WA 98052}$ , Alexei Kitaev$\footnote{Caltech, 1200 East California Boulevard, Pasadena, CA 91125}$ and Jacob Lurie$\footnote{MIT, 77 Massachusetts Avenue, Cambridge, MA 02139-4307}$}

\maketitle

\begin{abstract} Let $G$ be a compact connected Lie group with trivial center. Using the action of $G$ on its Lie algebra, we define an operator norm $|\,\,|_{G}$ which induces a bi-invariant metric $d_G(x,y)=|Ad(yx^{-1})|_{G}$ on $G$. We prove the existence of a constant $\beta \approx .12$ (independent of $G$) such that for any closed subgroup $H \subsetneq G$, the diameter of the quotient $G/H$ (in the induced
metric) is $\geq \beta$. \end{abstract}

\section{Introduction}
Finding a lower bound to the (operator norm) diameter of
homogeneous spaces $G/H$, $G$ compact is a natural geometric
problem.  It can also be motivated by considering quantum
computation.  In standard models [NC] the state space of a
(theoretical) quantum computer is a Hilbert space with a tensor
decomposition, $(\mathbb{C}^2 )^{\otimes n}$.  A $\lq\lq$gate" is
a local unitary operation acting on a small number, perhaps two,
tensor factors (and as the identity on the remaining factors). One
often wonders if a certain set of local gates is
$\lq\lq$universal" meaning that the closed subgroup $H$ they
generate satisfies $U(1) H= U(2^n )$. We produce a constant $\beta
\approx .12$ so that diam $U(2^n)/U(1)H < \beta$ implies
universality, where diameter is to be computed in the operator
norm. This norm is well-suited here because it is stable under
$\otimes_{\textnormal{id}}$.

Because the operator norm is bi-invariant it suffices to check
that every element $b$ in the ball of radius $2\beta$ about the
identity of $SU(2^n )$ has Ball$_\beta (b) \cap H \neq \emptyset$.
In principle this leads to an algorithm to test if a gate set is
universal.  Such an algorithm will be exponentially slow in n. But
often it is assumed that identical gates can be applied on any pair of
$\C^2$ factors; in this case universality for $n=2$ is sufficient
to imply universality for all $n$.

Let $G$ be a compact Lie group with trivial center.
The semisimplicity of $G$ implies that the (negative of the) Killing form is a natural positive-definite, bi-invariant inner product on the Lie algebra $\g$ of $G$. We let $||x||_{\g}$ denote the induced (Euclidean) norm on $\g$. We use this to define the {\it operator norm} on $G$ as follows: $$|g|_{G} = \sup_{||y||_{\g} = 1} |\angle(y, \Ad_{g} y)|$$ where $\angle(y, \Ad_{g} y)$ denotes the usual Euclidean angle between the vectors $y$ and $\Ad_{g} y$, normalized so that it lies in the interval $[-\pi,\pi]$. Since angles between vectors in a Euclidean space obey a triangle inequality, we deduce the inequality $|gh|_{G} \leq |g|_{G} + |h|_{G}$. It is also clear that $|g|_{G} = 0$ if and only if $\Ad_{g}$ is the identity, which implies that $g$ is the identity since the adjoint action of $G$ is faithful up to the center of $G$, and we have assumed that the center of $G$ is trivial.

We define a distance on $G$ by the formula $d_G(g,g') =
|g^{-1} g'|_{G}$. It is easy to check that this defines
a bi-invariant metric on $G$, where all distances are
bounded above by $\pi$. Note that $d_G$ is continuous
on $G$, hence there is a continuous bijection from $G$
with its usual topology to $G$ with the topology
induced by $d_G$. Since the source is compact and the
target Hausdorff (this fails if $G$ has nontrivial
center, since the operator norm of a central element is
equal to zero), we deduce that the metric $d_G$
determines the usual topology on $G$.

For any closed subgroup $H$ of $G$, the homogeneous
space $G/H$ inherits a quotient metric given by the
formula $$d_{G/H}(p,q) = \inf d_G(\widetilde{p},
\widetilde{q}) = \inf_{gp=q} |g|_{G}$$ where the first
infimum is taken over all pairs $\widetilde{p},
\widetilde{q} \in G$ lifting the pair $p,q \in G/H$.
Note that if $H$ is contained in $H'$, then the
diameter
of $G/H$ is at least as large as that of $G/H'$.

We are now in a position to state the main result:

\begin{theorem}\label{main} Let $G$ be a compact
connected Lie group with trivial center and
$H\varsubsetneq G$ a proper compact subgroup of $G$.
Then the diameter of $G/H$ with respect to the metric
$d_{G/H}$ is no smaller than $\beta$, where $\beta$ is
the smallest real solution to the transcendental
equation $\cos^2(\alpha - \beta) +
\sin^2(\alpha - \beta) \sin(\beta) = \cos(4
\beta)$ and $\cos(\alpha) = \frac{7}{8}$. \end{theorem}

One can estimate that the constant $\beta$ is approximately $.124332$.

\begin{example}\label{finalword}
Consider the case where $G = H \times H$ is a product, and $H$ is embedded diagonally. Choose an element $h \in H$ with $|h|_{H} = \pi$ (such an element exists in any nontrivial one parameter subgroup). Then in $H \times H$, the distance $d_{H \times H}( h \times 0, h' \times h')$ is equal to the larger of $d_H(h,h')$ and $d_H(h',e)$. By the triangle inequality, this distance is at least $\frac{\pi}{2}$. It follows that the diameter of $G/H$ is at least $\frac{\pi}{2}$. \end{example}

\noindent{\bf Remarks:}
\begin{description}
  \item[(1)] For any orthogonal representation $\tau: G \rightarrow O(V)$
  of a group $G$, we can define an operator norm on $G$ with respect to $V$:
  $$|g|_{G,\tau} = \sup_{||v||=1} | \angle(v, gv) | $$
  This construction has the following properties: \begin{itemize}
    \item If $V$ is the complex plane $\C$, and $g \in G$ acts by multiplication
    by $e^{i \alpha}$ where $- \pi \leq \alpha \leq \pi$, then $|g|_{G,\tau} = |\alpha|$.
    \item Given any subgroup $H \subseteq G$, the restriction of $|\,|_{G,\tau}$
    to $H$ is equal to $|\,|_{H,\tau|H}$.
    \item The operator norm associated to a direct sum of representations
    $\tau_i$ of $G$ is the supremum
    of the operator norms associated to the representations $\tau_i$.
    \item In particular, the operator norm on $G$ associated to a representation $V$ is
    identical with the operator norm on $G$ associated to the complexification
    $V \otimes_{\R} \C$ (with its induced Hermitian structure).
    \item To evaluate $|g|_{G,\tau}$, we can replace $G$ by the subgroup generated by $g$
    and $V$ by its complexification, which decomposes into one-dimensional complex eigenspaces
    under the action of $g$. We deduce that $|g|_{G,\tau}$ is the supremum of
    $| \log \lambda_j |$, where $\{\lambda_j\}$ is the set of eigenvalues for the action
    of $g$ on $V$ (and the logarithms are chosen to be of absolute value $\leq \pi$). \end{itemize}

  \item[(2)] The reader may be curious about the diameter of $G / H$ relative
  to the Riemannian quotient of the Killing metric $d_K$.
  If we let $N$ denote the dimension of ${\frak g}$, then we have
  \[ d \leq d_K \leq \frac{3 N^{\frac{1}{2}} d}{2} \]

  \item[(3)] We ask if the quotient $SO(3)/I$
 is the homogenous space of smallest diameter, where $I
\simeq A_5$ denotes the symmetry group of the icosahedron.

  \item[(4)] We wonder if there is a similar universal lower
  bound to the diameter of double coset spaces $K\diagdown G
  \diagup H$, $G$ as above, $K$, $H \subset G$ closed subgroups.  Our method does
  not apply directly.

  \item[(5)] Although suggested by a modern subject the theorem
  could easily have been proved a hundred years ago and in fact may
  have been (or may be) known.

\end{description}

\section{Small Subgroups}

Throughout this section, $G$ shall denote a compact, connected Lie group with trivial center. We give a quantitative version of the principle that discrete subgroups of $G$ generated by ``sufficiently small'' elements are automatically abelian. We will use this in the proof of Theorem \ref{main} in the case where $H$ is discrete.

We will need to understand the operator norm on $G$ a bit better. To this end, we introduce the {\it operator norm} $$|x|_{\g} = \sup_{||y||_{\g} = 1} || [x,y] ||_{\g}$$ on the Lie algebra $\g$ of $G$. This is a $G$-invariant function on $\g$, so we can unambiguously define the operator norm of any tangent vector to the manifold $G$ by transporting that tangent vector to the origin (via left or right translation) and then applying $x \mapsto |x|_{\g}$.

The operator norm on $\g$ is related to the operator norm on $G$ by the following:

\begin{lemma}\label{estimate}
The exponential map $x \mapsto \exp(x)$ induces a bijection between $\g_{0} = \{ x \in \g: |x|_{\g} < \frac{2 \pi}{3} \}$ and $G_0 = \{ g \in G: |g|_{G} < \frac{2 \pi}{3} \}$. This bijection preserves the operator norms. \end{lemma}

\begin{proof}
First, we claim that the map $x \mapsto \exp(x)$ does not increase the operator norm. This follows from the fact that the eigenvalues of $\exp(x)$ have the form $\exp(\kappa)$, where $\kappa$ is an eigenvalue of $x$. It follows that the exponential map sends $\g_0$ into $G_0$.

Choose $g \in G_0$, and fix a maximal torus $T$
containing $g$. Let ${\frak t}$ be the Lie algebra of
$T$. Decompose $\g \otimes_{\R} \C$ into eigenspaces
for the action of $T$: $\g \otimes_{\R} \C = {\frak t} \otimes_{\R} \C \oplus \bigoplus_{\alpha} \g_{\alpha}$. The element $g$ acts by an eigenvalue $\Lambda(\alpha)$ on each nonzero eigenspace $\g_{\alpha}$. Since $g$ is an orthogonal transformation, we may write $\Lambda(\alpha) = e^{i \lambda(\alpha)}$. Since $g \in G_0$, it is possible to choose the function $\lambda$ so that $- \frac{2 \pi}{3} < \lambda(\alpha) < \frac{2 \pi}{3}$ for each root $\alpha$. This determines the function $\lambda$ uniquely.

Choose a system $\Delta$ of simple roots, and let $x$ be the unique element of ${\frak t}$ such that $\alpha(x) = \lambda(\alpha)$ for each $\alpha \in \Delta$. It follows immediately that $\exp(x) = g$ (since $G$ has trivial center). To show that $x \in \g_0$, we need to show that $|\alpha(x)| < \frac{2 \pi}{3}$ for all roots $\alpha$. For this, it will suffice to prove that $\alpha(x) = \lambda(\alpha)$ for all roots $\alpha$.

The uniqueness of $\lambda$ implies immediately that $\lambda(-\alpha) = - \lambda(\alpha)$. Thus, it will suffice to prove that the equation $\alpha(x) = \lambda(\alpha)$ holds when $\alpha$ is positive (with respect to the root basis $\Delta$). Since the equation is known to hold whenever $\alpha \in \Delta$, it will suffice to prove that
$\alpha(x) = \lambda(\alpha)$, $\beta(x) = \lambda(\beta)$ implies
$$(\alpha+\beta)(x) = \lambda(\alpha+\beta).$$
In other words, we need to show that the quantity
$$\epsilon = \lambda(\alpha+\beta) - \lambda(\alpha) - \lambda(\beta)$$ is equal to zero. By construction, $|\epsilon| < 2 \pi$. On the other hand, since
$\Lambda(\alpha) \Lambda(\beta) = \Lambda(\alpha+ \beta)$, we deduce that $e^{i \epsilon} = 1$, so that $\epsilon$ is an integral multiple of $2 \pi$. It follows that $\epsilon = 0$, as desired.

It is clear from the construction that $|x|_{\g} = |g|_{G}$. To complete the proof, we need to show that $g$ has no other logarithms lying in $\g_0$. This follows from the fact that any unitary transformation (in particular, the adjoint action of $g$ on ${\frak g}$) which does not have $-1$ as an eigenvalue has a unique logarithm whose eigenvalues are of absolute value $< \pi$. \end{proof}

\begin{lemma}\label{est}
Let $p: [0,1] \rightarrow G$ be a smooth function with $p(0)$ equal to the identity of $G$. Then $|p(1)|_{G} \leq \int^1_0 |p'(t)|_{\g} dt$. \end{lemma}

\begin{proof}
For $N$ sufficiently large, we can write $p(\frac{i+1}{N}) = p(\frac{i}{N}) \exp( \frac{x_i}{N} )$, where $x_i$ is approximately equal to the derivative of $p$ at $\frac{i}{N}$. Thus, as $N$ goes to $\infty$, the average $\frac{ |x_0|_{\g} + \ldots + |x_{N-1}|_{\g} }{N}$ converges to the integral on the right hand side of the desired inequality. By the triangle inequality, it will suffice to prove that $|p(\frac{i}{N})^{-1} p(\frac{i+1}{N})|_G \leq |\frac{|x_i|_{\g}}{N}|_{\g}$. If $N$ is sufficiently large, then this follows immediately from Lemma \ref{estimate}. \end{proof}

\begin{remark} The metric $d_G$ on $G$ is not
necessarily a path metric: given $g,h \in G$, there
does not necessarily exist a path in $G$ having length
equal to $d_G(g,h)$. However, it follows from Lemma \ref{estimate} that $d_G$ is a path metric {\em locally on $G$}. The length of a (smooth) path can be obtained by integrating the operator norm of the derivative of a path. Replacing $d_G$ by the associated path metric only increases distances, so that Theorem \ref{main} remains valid for the path metric associated to $d_G$. This modified version of Theorem \ref{main} makes sense (and remains true) for compact Lie groups $G$ with finite center. \end{remark}

We can now proceed to the main result of this section.
Let $\alpha$ denote the smallest positive real number satisfying
$\cos(\alpha) = \frac{7}{8}$.

\begin{theorem}\label{good} Let $H \subset G$ be a discrete subgroup. Let $h,k \in H$ and suppose $|h|_G < \frac{\pi}{2}$, $|k|_G < \alpha$. Then $[h,k] = 1$. \end{theorem}

\begin{proof} We define a sequence of elements of $G$ by recursion as
follows: $h_0 = h$, $h_{n+1} = [h_{n},k]$. Let $C$ satisfy the equation $\frac{C^2}{4} = 2 - 2 \cos |k|_{G}$. Then the assumption on $k$ ensures that $C < 1$. Our first goal is to prove that the operator norm of the sequence $\{h_{n}\}$ obeys the estimate $|h_{n}|_G < C^n \frac{\pi}{2}$. For $n=0$, this is part of our hypothesis. Assuming that the estimate $|h_{n}|_{G} < C^{n} \frac{\pi}{2}$ is valid, we can use Lemma \ref{estimate} to write $h_{n} = \exp(x)$, $|x|_{\g} < C^{n} \frac{\pi}{2}$. Now define $p(t) = [ \exp(tx), k ]$, so that $p(0) = 1$ and $p(t) = h_{n+1}$.

Using Lemma \ref{est}, we deduce that $|h_{n+1}|_{G} \leq \int_0^1
|p'(t)|_{\g} \leq \sup_{t} |p'(t)|_{\g}$. On the other hand, the vector
$p'(t)$ can be written as a difference $$R_{p(t)} x - L_{ \exp(tx) k \exp(-tx)} R_{k^{-1}} x$$ where $R_{g}$ and $L_g$ denote left and right translation by $g$. We obtain $$\begin{array}{ccl} |p'(t)|_{\g} & = & | x
- \Ad_{ \exp(tx) k \exp(-tx)} x |_{\g} \\ & = & | \Ad_{\exp(-tx)} x - \Ad_{k \exp(-tx)} x |_{\g} \\ & = & | x - \Ad_{k} x |_{\g} \\ & = & \sup_{||y||_{\g} = 1} || [x - \Ad_{k} x, y] ||_{\g} \\ & \leq & \sup_{||y||_{\g} = 1} (||[x,y] - Ad_{k} [x,y]||_{\g}
+ || Ad_{k}[x,y] - [Ad_{k} x, y] ||_{\g}) \\
& \leq & \sup_{||y||_{\g} = 1} ||[x,y] - Ad_{k} [x,y]||_{\g}
+ \sup_{||y||_{\g} = 1} || [x, y - Ad_{k}^{-1} y] ||_{\g} \\
& \leq & \sqrt{2 -2 \cos |k|_G} \sup_{||y||_{\g}=1} ||[x,y]||_{\g}
+ |x|_{\g} \sup_{||y||_{\g}=1} ||y-Ad_{k}^{-1} y] ||_{\g} \\
& \leq & 2 \sqrt{2 - \cos |k|_G} |x|_{\g} \\
& = & C |x|_{\g} \\ & <
& C^{n+1} \frac{\pi}{2}, \end{array} $$ as desired.

It follows that the operator norms of the sequence $\{h_{n}\}$ converge to zero. Therefore the sequence $\{h_{n} \}$ converges to the identity of $G$. Since $H$ is a discrete subgroup, it follows that $h_{n}$ is equal to the identity if $n$ is sufficiently large. We will next show that $h_{n} = 1$ for all $n > 0$, using an argument of Frobenius which proceeds by a descending induction on $n$. Once we know that $h_1 = 1$, the proof will be complete.

Assume that $h_{n+1} = 1$. Then $k$ commutes with $h_{n}$, and therefore also with $h_{n}k = h_{n-1} k h_{n-1}^{-1}$. It follows that $\g \otimes_{\R} \C$ admits a basis whose elements are eigenvectors for both $k$ and $h_{n-1} k h_{n-1}^{-1}$. If the eigenvalues are the same in both cases, then we deduce that $k = h_{n-1} k h_{n-1}^{-1}$, so that $h_{n}$ is the identity and we are done. Otherwise, there exists $v \in \g \otimes_{\R} \C$ which is an eigenvector for both $k$ and $h_{n-1} k h_{n-1}^{-1}$, with different eigenvalues. Equivalently, both $v$ and $h_{n-1} v$ are eigenvectors for $k$, with different eigenvalues. Thus $v$ and $h_{n-1} v$ are orthogonal, which implies $|h_{n-1}|_{G} \geq \frac{\pi}{2}$, a contradiction. \end{proof}

\section{The Proof when $H$ is Discrete}

In this section, we will give the proof of Theorem \ref{main} in the case where $H$ is a discrete subgroup. The idea is to show that if $G/H$ is too small, then $H$ contains noncommuting elements which are close to the identity, contradicting Theorem \ref{good}.

In the statements that follow, we let $\alpha$ denote the smallest positive real solution to $\cos(\alpha) = \frac{7}{8}$ and $\beta$ the smallest positive real solution to the transcendental equation $\cos^2(\alpha - \beta) + \sin^2(\alpha - \beta) \cos(\frac{\pi}{2} - \beta) = \cos(4 \beta)$.

\begin{lemma}\label{goodd} Let $G$ be a compact, connected Lie group with trivial center. Then there exist elements $h,k \in G$ having the property that for any $h',k' \in G$ with $d_G(h,h'), d_G(k,k') < \beta$, we have $|h'|_{G} < \frac{\pi}{2}$, $|k'|_{G} < \alpha$, and $[h',k'] \neq 1$. \end{lemma}

\begin{proof} Choose a (local) embedding $p: SU(2) \rightarrow G$ corresponding to a root of some simple component of $G$. We will assume that if the relevant component has roots of two different lengths, then the embedding $p$ corresponds to a long root. This ensures that the weights of $SU(2)$ acting on $\g$ are no larger than the weights of the adjoint representation.

In the Lie algebra $\so(3)$ of $SU(2)$, we let $x$ and $y$ denote infinitesimal rotations of angles $\frac{\pi}{2} - \beta$ and $\alpha - \beta$ about orthogonal axes. Then, by the above condition on weights, we deduce that $h=p(\exp(x))$ and $k=p(\exp(y))$ satisfy the conditions $|h|_{G} = \frac{\pi}{2} - \beta$, $|k|_{G} = \alpha - \beta$.

We claim that the pair $h,k \in G$ satisfies the conclusion of the lemma. To see this, choose any pair $h',k' \in G$ with $d(h,h'), d(k,k') < \beta$. Then we deduce $|h'|_{G} < \frac{\pi}{2}$, $|k'|_{G} < \alpha$ from the triangle inequality. To complete the proof, we must show that $h'$ and $k'$ do not commute. To see this, we let $v$ denote the image in $\g$ of a vector in $\so(3)$ about which $x$ is an infinitesimal rotation. Then $hv = v$, while $\angle(v,kv) = \alpha
- \beta$. Elementary trigonometry now yields $$\begin{array}{ccl}
\angle(hkv,khv) &= & \angle(hkv,kv) \\ &= & \cos^{-1} ( \cos^2 ( \alpha - \beta) + \sin^2( \alpha - \beta ) \cos( \frac{\pi}{2} - \beta ) )\\ & = & \cos^{-1} ( \cos(4 \beta) ) = 4 \beta.\\ \end{array} $$ By the triangle inequality, we get $$ \begin{array}{ccl} 4 \beta &=& \angle(hkv, khv) \\ & \leq & \angle(hkv,h'kv) + \angle (h'kv,
h'k'v) + \angle( h'k'v, k'h'v) \\ & & + \angle( k'h'v, k'hv) + \angle(k'hv, khv) \\ & < & 4 \beta + \angle(h'k'v, k'h' v),\\ \end{array} $$ which implies $\angle(h'k'v, k'h' v) > 0$ so that $[h',k'] \neq 1$. \end{proof}

We can now complete the proof of Theorem \ref{main} in the case where $H$ is discrete:

\begin{proof}
Choose $h,k \in G$ satisfying the conclusion of Lemma \ref{goodd}. Since $G/H$ has diameter less than $\beta$, the cosets $hH$ and $kH$ are within $\beta$ of the identity coset in $G/H$, which implies that there exist $h',k' \in H$ with $d(h,h'), d(k,k') < \beta$. Lemma \ref{goodd} ensures that $h'$ and $k'$ do not commute, which contradicts Theorem \ref{good}. \end{proof}

\section{The Proof when $G$ is Simple}

In this section, we give the proof of the main theorem in the case where $H$ is nondiscrete and $G$ is simple. The idea in this case is to show that because the Lie algebra ${\frak h}$ of $H$ cannot be a $G$-invariant subspace of $\g$, the action of $G$ automatically moves it quite a bit: this is made precise by Theorem \ref{excellent}. Since ${\frak h}$ is invariant under the action of $H$, this will force $G/H$ to have large diameter in the operator norm.

We begin with some general remarks about angles between subspaces of a Hilbert space. Let $V$ be a real Hilbert space, and let $U,W \subseteq V$ be linear subspaces. The angle $\angle(U,W)$ between $U$ and $W$ is defined to be $$\max(\sup_{u \in U - \{0\}} \inf_{w \in W - \{0\}} |\angle(u,w)|, \sup_{w \in W - \{0\}} \inf_{u \in U - \{0\}} |\angle(u,w))|).$$

Note that for a fixed unit vector $u \in U$, the cosine of the minimal angle $\angle(u,w)$ with $w \in W$ is equal to the length of the orthogonal projection of $u$ onto $W^{\perp}$. Thus, the sine of the minimal (positive) angle is equal to the length of the orthogonal projection of $u$ onto ${W}^{\perp}$. Consequently we have $$\sin(\sup_{u \in  U - \{0\}} \inf_{w \in W - \{0\}} |\angle(u,w)| \,)= \sup_{||u|| = 1, ||{w}^{\perp}|| = 1} \leftbracket u, {w}^{\perp} \rightbracket$$ which is symmetric in $U$ and ${W}^{\perp}$. From this symmetry we can deduce:

\begin{lemma}
For any pair of subspaces
$U,W \subseteq V$, the angle $\angle(U,W)$
is equal to the angle $\angle(U^{\perp}, {W}^{\perp})$. \end{lemma}

We will also need the following elementary fact:

\begin{lemma}\label{trace}
Let $V$ be a finite-dimensional Hilbert space, and let $A$ be an endomorphism of $V$ having rank $k$. Then $|\trace(A)| \leq k |A|$. \end{lemma}

\begin{proof}
Choose an orthonormal basis $\{v_i\}_{1 \leq i \leq n}$
for $V$ having the property that $A v_i = 0$ for $i > k$. Then $$|\trace(A)| = | \sum_{i} \leftbracket v_i, A v_i \rightbracket | \leq \sum_{1 \leq i \leq k} | \leftbracket v_i, A v_i \rightbracket | \leq \sum_{1 \leq i \leq k} |A| = k |A|$$ \end{proof}

We now proceed to the main point.

\begin{theorem}\label{excellent} Let $G$ be a compact Lie group acting irreducibly on a (necessarily finite dimensional) complex Hilbert space $V$. Let $W \neq 0,V$ be a nontrivial subspace. Then there exists $g \in G$ such that $\angle(W,gW) \geq \frac{\pi}{4}$. \end{theorem}

\begin{proof} Suppose, to the contrary, that $\angle(W,gW) < \frac{\pi}{4}$ for all $g \in G$. Let $V$ have dimension $n$. Replacing $W$ by $W^{\perp}$ if necessary, we may assume that the dimension $k$ of $W$ satisfies $k \leq \frac{n}{2}$.For any subspace $U \subseteq V$, we let $\Pi_U$ denote the orthogonal projection onto $U$.

For each $g \in G$, projection from $gW$ onto $W^{\perp}$ or from $W^{\perp}$ to $gW$ shrinks lengths by a factor of $\sin \angle(W,gW) \leq \sin \frac{\pi}{4}$ at least. It follows that $$| \Pi_{W^{\perp}} \Pi_{gW} \Pi_{W^{\perp}} | \leq | \Pi_{W^{\perp}} \Pi_{gW}| \,\, |\Pi_{gW} \Pi_{W^{\perp}}| < \frac{1}{2}.$$

Using the identity $\trace(AB) = \trace(BA)$, we deduce $$\begin{array}{ccl} \trace(\Pi_{gW} \Pi_{W^{\perp}}) & = & \trace( \Pi_{gW} \Pi_{W^{\perp}} \Pi_{W^{\perp}}) \\ & = & \trace(\Pi_{W^{\perp}} \Pi_{gW} \Pi_{W^{\perp}}) \leq k | \Pi_{W^{\perp}} \Pi_{gW} \Pi_{W^{\perp}} | \\ & < & \frac{k}{2}.\\ \end{array}$$

Integrating this result over $G$ (with respect to a Haar measure which is normalized so that $\int_{G} 1 = 1$), we deduce $$\trace( (\int_G
\Pi_{gW}) \Pi_{W^{\perp}} ) = \int_G \trace ( \Pi_{gW} \Pi_{W^{\perp}} ) < \frac{n}{2}.$$

On the other hand, $\int_G \Pi_{gW}$ is a $G$-invariant element of $\End(V)$. Since $V$ is irreducible, Schur's lemma implies that $\int_G \Pi_{gW} = \lambda 1_{V}$ for some scalar $\lambda \in \C$. We can compute $\lambda$ by taking traces: $$ \begin{array}{ccl} n \lambda &=& \trace( \lambda 1_{V}) \\ &= &\trace ( \int_{G} \Pi_{gW} ) \\ &=& \int_{G}
\trace(\Pi_{gW}) = k,\\ \end{array}$$ so that $\lambda = \frac{k}{n}$. Thus $\frac{k(n-k)}{n} = \trace( \frac{k}{n} \Pi_{W}^{\perp}) < \frac{k}{2}$, so that $2(n-k) < n$, a contradiction. \end{proof}

From Theorem \ref{excellent}, one can easily deduce the analogous result in the case when $V$ is a real Hilbert space, provided that $V \otimes_{\R} \C$ remains an irreducible representation of $G$. Using this, we can easily complete the proof of Theorem \ref{main} in the case where $G$ is simple and $H$ is nondiscrete (with an even better constant).

\begin{proof}
Let ${\frak h}$ denote the Lie algebra of $H$. Since $H \neq G$ and $G$ is connected, $\frak h \subsetneq \g$. Since $H$ is nondiscrete, ${\frak h} \neq 0$. Since $\g \otimes_{\R} \C$ is an irreducible representation of $G$, we deduce that there exists $g \in G$ such that $\angle(g {\frak h}, {\frak h}) \geq \frac{\pi}{4}$. Now one deduces that for any $h \in H$, $gh' \in gH$, the distance $$d(gh', h) = | gh' h^{-1} |_G \geq \angle( g h' h^{-1} {\frak h}, {\frak h}) = \angle( g {\frak h}, {\frak h}) \geq \frac{\pi}{4}.$$ It follows that the distance between the cosets $gH$ and $H$ in $G/H$ is at least $\frac{\pi}{4}$. \end{proof}

\section{The General Case}

We now know that Theorem \ref{main} is valid under the additional assumption that the group $G$ is simple. We will complete the proof by showing how to reduce to this case. The main tool is the following observation:

\begin{proposition}\label{used}
Let $\pi: G \rightarrow G'$ be a surjection of compact connected Lie groups with trivial center, let $H$ be a closed subgroup of $G$ and $H' = \pi(H)$ its image in $G'$. Then
$\diam(G'/H') \leq \diam(G/H)$.
\end{proposition}

\begin{proof} For any points $x',y' \in G'/H'$, we can lift them to a pair of points $x,y \in G/H$. It will suffice to show $d_{G/H}(x,y) \geq d_{G'/H'}(x',y')$. The left hand side is equal to $$\inf_{g x =y}
|g|_{G}$$ and the right hand side to $$\inf_{g' x' = y'} |g'|_{G'}.$$ To
complete the proof, it suffices to show that $|g|_{G} \geq |\pi(g)|_{G'}$. This follows immediately since we may identify the Lie algebra $\g'$ of $G'$ with a direct summand of $\g$. \end{proof}

Now assume that $G$ is a compact, connected Lie group with trivial center. Then it is a product of simple factors $\{ G_{\alpha} \}_{\alpha \in \Lambda}$. Let $\pi_{\alpha}: G \rightarrow G_{\alpha}$ denote the projection. Let $H \subsetneq G$ be a closed subgroup. If $\pi_{\alpha} H \neq G_{\alpha}$ for some $\alpha \in \Lambda$, then $\diam(G/H) \geq \diam(G_{\alpha}/ \pi_{\alpha} H) \geq \beta$ and we are done. Otherwise, $\pi_{\alpha}$ induces a surjection of Lie algebras ${\frak h} \rightarrow {\frak g}_{\alpha}$ for each $\alpha$. By the structure theory of reductive Lie algebras, we deduce that ${\frak h} = {\frak h}_{\alpha} \oplus {\frak k}_{\alpha}$, where $\pi_{\alpha}$ is zero on ${\frak k}_{\alpha}$ and induces an isomorphism ${\frak h}_{\alpha} \simeq {\frak g}_{\alpha}$. Since ${\frak h}_{\alpha}$ is therefore simple, ${\frak k}_{\alpha}$ may be characterized as the centralizer of ${\frak h}_{\alpha}$ in ${\frak h}$.

Since $H \neq G$ and $G$ is connected, $H$ must have smaller dimension than $G$. It follows that the subalgebras ${\frak h}_{\alpha} \subseteq {\frak h}$ cannot all be distinct. Choose $\alpha, \alpha' \in \Lambda$ with ${\frak h}_{\alpha} = {\frak h}_{\alpha'}$. The the map $H \rightarrow G_{\alpha} \times G_{\alpha'}$ is not surjective on Lie algebras. Without loss of generality, we may replace $G$ by $G_{\alpha} \times G_{\alpha'}$ and $H$ by its image in $G_{\alpha} \times G_{\alpha'}$.

Since the Lie algebra of $H$ now maps isomorphically onto the Lie algebras of the factors $G_{\alpha}$ and $G_{\alpha'}$, it follows that the connected component $H_0$ of the identity in $H$ is isomorphic to $G_{\alpha}$, which is included diagonally in $G_{\alpha} \times G_{\alpha'}$. Then $H = H_0 (H \cap (G_{\alpha} \times 1))$. The intersection $K=H \cap (G_{\alpha} \times 1)$ is normalized by $H_0 = \{(g,g): g \in G_{\alpha} \}$, hence it is normalized by $G_{\alpha} \times \{e\}$. Since $G_{\alpha'}$ is simple, we deduce that $K=\{e\}$. Thus $H=H_0$ is embedded diagonally in $G_{\alpha} \times G_{\alpha'}$. We have already considered this case in Example \ref{finalword}, where we saw that the diameter of $G'/H'$ is at least $\frac{\pi}{2}$.

\begin{remark}
If we restrict our attention to the case where $H$ is a {\em connected} subgroup of $G$, then our proof gives a better lower bound of $\frac{\pi}{4}$. \end{remark}

\end{document}